\documentclass[aps,prb,preprint]{revtex4}
\usepackage{graphicx}
\usepackage{ulem}

\begin{document}

\baselineskip=1.5cm

\title{Correlations between morphology, crystal structure and magnetization of epitaxial cobalt-platinum films grown with pulsed laser ablation}

\author{R. K. Rakshit, S. K. Bose, R. Sharma and R. C. Budhani$^{a}$}

\affiliation{Condensed Matter - Low
Dimensional Systems Laboratory, Department of Physics, Indian
Institute of Technology Kanpur, Kanpur - 208 016, India}
\author{T. Vijaykumar, S. J. Neena and G. U. Kulkarni}
\affiliation{Jawaharlal Nehru Centre for Advanced Scientific
Research, Jakkur P.O., Bangalore 560 064, India}

\begin{abstract}
\baselineskip=1.5cm The effects of growth rate (G$_r$), deposition
temperature (T$_d$), film thickness (t$_F$), and substrate induced
strain ($\epsilon$) on morphological, crystallographic and
magnetic characteristics of equiatomic CoPt epitaxial films
synthesized with pulsed laser deposition (PLD) are investigated.
The (001) oriented single crystal substrates of MgO, SrTiO$_3$ and
LaAlO$_3$ provide different degree of epitaxial strain for growth
of the disordered face centered cubic (fcc) and ordered face
centered tetragonal (L1$_0$) phases of  CoPt.  The films deposited
at T$_d$ $\approx$ 600 $^0$C on all three substrates are fcc with
in-plane magnetization and a narrow hysteresis loop of width
$\approx$ 200 Oe. The L1$_0$ phase, stabilized only at T$_d$
$\geq$ 700 $^0$C becomes predominantly c-axis oriented as T$_d$ is
increased to 800 $^0$C. While the crystallographic structure of
the films depends solely on the T$_d$, their microstructure and
magnetization characteristics are decided by the growth rate. At
the higher G$_r$ ($\approx$ 1 {\AA}/sec) the L1$_0$ films have a
maze-like structure which converts to a continuous film as the
t$_F$ is increased from 20 nm to 50 nm. The magnetic coercivity of
these films increases as the L1$_0$ phase fraction grows with
T$_d$ and its orientation becomes out of the film plane. The
evolution of microstructure with T$_d$ is remarkably different at
lower growth rate ($\approx$ 0.4 {\AA}/sec). Here the structure
changes from a self-similar fractal pattern to a disordered
assembly of nano-dots as the T$_d$ is raised from 700 to 800
$^0$C, and is understood in terms of the imbalance between strain
and interfacial energies. Magnetic force microscopy of such films
reveals no distinct domain walls within the nano-islands while a
clear contrast is seen between the islands of reversed
magnetization. Magnetic relaxation measurements on these
assemblies of single-domain islands show negligible decay of
magnetization unless a reverse field close to the coercive field
(H$_c$ $\approx$ 30 kOe) is applied. The simple picture of
coherent rotation of moment appears incompatible with the time
dependence of the remanent magnetization in these films.

\footnotetext[1]{Author to whom correspondence should be
addressed; electronic mail: rcb@iitk.ac.in}

\end{abstract}


\maketitle
\section{Introduction}
Studies of the correlations between crystallographic as well as
morphological structure and magnetic coercivity of CoPt and FePt
alloy films have been of considerable interest in recent years due
to the strong perpendicular magnetic anisotropy and chemical
inertness of these alloys which make them suitable for high
density recording applications
\cite{Shima,Liou,Aoyama,Ristau,Jeong,Weller,Kim,Park,Turpanov,Farrow,Ersen,Xiao,Hao,Lewis,Castaldi,
Rakshit,Budhani}. The key structural feature responsible for
magnetic anisotropy of these systems is the ordering of Pt and Co
or Fe atoms in the crystal lattice. The ordered L1$_0$ phase at
equiatomic composition comprises of an alternate sequence of Co
(Fe) and Pt planes along the crystallographic c-axis
\cite{Newkirk}. Studies of coercivity and magnetization reversal
in granular thin films of CoPt and FePt are of interest as well.
When the size of grains becomes small enough to make them
mono-domain, the magnetic response of the system may differ
significantly from that in the bulk form. Studies of thermal
stability of magnetization and the process of its reversal are
critical for high density recording. The reversal can occur either
continuously through coherent or incoherent rotation of spins, or
discontinuously through nucleation of reversed domains and their
expansion through domain wall displacement \cite{Stoner,
Livingston}. Growth of reverse domains can be impeded by defects,
such as second-phase precipitates or grain boundaries which pin
the domain walls by locally altering their energy. In epitaxial
films consisting of discontinuous islands/dots of sufficiently
small size such that domain wall formation is energetically
unfavorable, the demagnetization process can have several
interesting features including quantum tunneling at sufficiently
low temperatures.

The experimental conditions that drive a particular growth
mechanism of CoPt (FePt) films and the linkage between
crystallographic and morphological structure, vis-a-vis
magnetocrystalline anisotropy, saturation and remanent
magnetization, coercive field and magnetic domain structure need
to be studied for each process of growth. Most of the earlier
research on thin films of CoPt employed sputtering \cite{Shima,
Ristau, Liou, Jeong, Kim, Turpanov} or molecular beam epitaxy
(MBE) \cite{Farrow, Ersen} to grow textured and epitaxial films.
The majority of these works have been carried out on (001) MgO
substrates. The lattice parameter of MgO (cubic, a = 4.21 {\AA})
is larger by 8.5 and 9.6 $\%$ compared to the basal plane unit
cell parameter of L1$_{0}$ FePt and CoPt respectively. This large
lattice mismatch invariably affects epitaxial growth and magnetic
properties of the films, whose extent needs to be quantified by
depositing fims on single crystal substrates which provide
different degree of lattice mismatch.

One of the promising thin film growth techniques that has been
used rarely for deposition of epitaxial CoPt is Pulsed Laser
Deposition (PLD) \cite{Budhani}. One of the key advantages of PLD
is a precise control over the stoichiometry of a multi-elemental
film. This has been demonstrated amply in perovskite oxide films
such as those of high T$_c$ superconductors and hole-doped
manganites. Additionally, the high adatom energy, which can be as
large as $\simeq$ 400 eV in PLD can assist in stabilizing
non-equilibrium phases \cite{Chrisey2}. Here we report PLD growth
of epitaxial CoPt films on (001) cut LaAlO$_3$ (LAO), SrTiO$_3$
(STO) and MgO single crystals with the objectives to; (i)
demonstrate the feasibility of growth of stoichiometric films from
ablation of an alloy target, (ii) vary the extent of substrate
induced strain by depositing films on single crystal substrates of
different lattice parameter such that effects of both tensile as
well compressive strain on growth morphology are addressed, (iii)
establish correlations between film morphology, phase purity and
magnetic properties, (iv) determine minimum growth temperature to
stabilize the high coercive tetragonal phase, (v) address the role
of kinetic conditions, such as deposition rate, in stabilizing a
particular growth morphology at several growth temperatures and
(vi), understand the demagnetization process in films of different
microstructures.

We note that the surface morphology of these films depends
strongly on deposition temperature (T$_d$) and growth rate
(G$_r$), in addition to the lattice mismatch with the substrate,
which can be characterized in terms of the strain parameter
$\epsilon$ defined as $ \epsilon {\rm{ }} = {\rm{ }}{{a_{bulk}  -
a_{substrate} } \over {a_{bulk} }}{\rm{ }} \times {\rm{ }}100 $,
where a is the lattice parameter. Values of the parameter
$\epsilon$ for in-plane and out-of-plane c-axis growth of the
L1$_0$ phase, and for the fcc phase on three different substrates
used in this study are listed in Table - I. For MgO (001) [a =
4.21 {\AA}], the $\epsilon$ for out-of-plane and in-plane c-axis
growth of the L1$_0$ phase is - 10.7 $\%$ and - 13.8 $\%$
respectively. The films on STO (001) [a = 3.905 {\AA}] also
experience tensile strain in both orientations but it is nominal
($\epsilon$ = - 2.7 $\%$ for out-of-plane and - 5.5 $\%$ for
in-plane c-axis growth) as compared to that on MgO (001). The
(001) face of LAO [cubic, a = 3.792 {\AA}], however provides an
ideal square planar lattice with minimum lattice mismatch (- 2.5
and 0.3 $\%$ for in-plane and out-of-plane c-axis growth
respectively) for epitaxial growth of the L1$_{0}$ phase of CoPt
in either orientation. The choice of different $\epsilon$ allows
tuning of the strain and hence the surface morphology of the
films, provided the flux of adatoms on the substrate is kept low.
At high flux densities, kinetic considerations eliminate the
morphological features evolved due to an imbalance of strain and
interfacial energies of the film. The best c-axis oriented
epitaxial L1$_{0}$ CoPt thin films display a coercive field
(H$_{c\perp}$) as high as $\approx$ 30 kOe and anisotropy energy
density of $\sim$ 4.62 $\times$ 10$^6$ J/m$^3$ at ambient
temperature.

\section{Experimental procedure}
Cobalt-platinum films of equiatomic composition were grown by
pulsed laser ablation of CoPt alloy target which was synthesized
by arc melting of high purity (4N) cobalt and platinum chunks
followed by homogenization at high temperature (900 $^0$C). A KrF
excimer laser (wavelength $\lambda$ = 248 nm and pulse width
$\sim$ 20 nsec) was used to ablate the 1 cm diameter target in an
all-metal-seal PLD chamber maintained at 50 mTorr pressure of
ultra high purity (99.9999 $\%$) nitrogen. Three 3 $\times$ 3
mm$^{2}$ pieces, one each of MgO, STO and LAO, all (001) cut, were
glued with silver epoxy on a substrate heater specially designed
for ultra high vacuum applications. The buffer gas (N$_2$), which
helps in reducing particulate density on the film surface, was
pumped out on completion of growth to ensure that no stable
nitrides of cobalt are formed when the film is cooled from the
deposition temperature (T$_d$ $\geq$ 600 $^0$C).

Three series of samples have been prepared under different growth
conditions as listed in Table - II, to evaluate their effect on
morphology, crystallographic structure and magnetic properties of
the films. The purpose of preparing each series is as follows; In
series `A' samples of t$_F$ $\approx$ 100 nm grown on (001) STO at
600, 700 and 800 $^0$C the effects of T$_d$ and post deposition
annealing were studied. Series `B' films were prepared to study
the effect of film thickness. This series consists of four films
of t$_F$ ranging from 20 to 100 nm grown at 700 $^0$C on STO and
MgO substrates followed by annealing for 30 minutes at the growth
temperature. The growth rate for these two sets of samples was
$\approx$ 1 {\AA}/sec. The last series, series `C', was prepared
at a slower growth rate ($\approx$ 0.4 {\AA}/sec). It consists of
three set of samples deposited at 700, 750 and 800 $^0$C on LAO,
STO and MgO mounted side-by-side on the substrate heater. The
thickness of these films was $\simeq$ 50 nm. All thicknesses
mentioned above were measured with a stylus based surface
profilometer.

X-ray diffraction (XRD) measurements were performed using the
Seifert model 3000P diffractometer with CuK$_{\alpha1,\alpha2}$
radiation in the standard $\theta$ - 2$\theta$ geometry with
scattering vector normal to the plane of the film. The surface
topography of the films was characterized using Scanning Electron
Microscopy (SEM). The energy dispersive X-ray element analyzer of
the same SEM was employed to ascertain the Co / Pt ratio in the
ablation target as well as in films. Magnetic Force Microscopy
(MFM) and Atomic Force Microscopy (AFM) measurements were
performed simultaneously on thermally demagnetized samples. A
multi-mode AFM with a Nanoscope IV controller (Digital
Instruments, Santa Barbara) was used in the tapping mode at a lift
height of 50 nm to probe the magnetic domain structure in films.
The MFM tip (CoCr coated Si) was magnetized vertically along the
tip axis, thereby allowing detection of the perpendicular
component of stray field emanating from the sample surface with a
spatial resolution of $\approx$ 50 nm.

Measurements of magnetization and magnetic relaxation were carried
out using a superconducting quantum interference device (SQUID)
based magnetometer [Quantum Design MPMS 5XC]. Both in-plane and
out-of-plane field orientations were used in these measurements
and all data were corrected for the contribution of the substrate.
The absolute value of moment has an error of $\pm$ 15 $\%$ due to
the uncertainty in the measurement of sample volume. For the
measurements of magnetic relaxation, sample was first magnetized
at 300 K with a 46 kOe field applied along the easy axis, and then
the moment was monitored as a function of time in a reversed field
of strength varying from 0 to $\leq$ H$_c$.

\section{Results and discussion}
\subsection{Effects of growth temperature and post growth annealing:}
In Fig. 1(a) we show the XRD profiles of the series `A' samples
(deposited on STO at 600, 700 and 800 $^0$C) together with the
diffraction profile of a bare substrate. The pattern for the
sample deposited at 600 $^0$C shows only one peak at 2$\theta$ =
48.12$^0$, which can be attributed to the (200) reflection of the
disordered A1 (fcc) phase. This reflection developed into a
doublet on increasing the T$_d$ to 700 $^0$C, and finally goes
back to a single peak shifted to higher 2$\theta$ at T$_d$ = 800
$^0$C. The doublet seen here is due to the mixing of the (200) and
(002) reflections of the L1$_0$ phase with some c-axis grains
oriented parallel and others oriented perpendicular to the film
plane. At the highest deposition temperature, i.e., 800 $^0$C,
although the (002) peak of the L1$_0$ phase becomes prominent, a
discernible shoulder remains on its low angle side. A marked
increase in the intensity and sharpness of (001), (002) and (003)
peaks of this film indicates significant improvement in the degree
of chemical ordering at the elevated T$_d$.

Fig. 1(b) presents the in-plane magnetization (M-H$_{\parallel}$)
from + H$_{max}$ to - H$_{max}$ of the films whose X-ray
diffraction data are shown in Fig. 1(a). Only half loops were
taken to save the SQUID time. Since there are three ordered
variants of CoPt which may co-exist in the film, each with c-axis
parallel to one of the cube edge directions i.e., [001], [010] and
[100] of the substrate [see, inset Fig. 1(b)], the external field
was applied parallel to the film plane at 45$^0$ with respect to
the [100] and [010] direction of the substrate as shown in the
lower inset of Fig. 1(b). The coercive field (H$_{c\parallel}$)
deduced from these measurements is 104 Oe, 6 kOe and 8.4 kOe for
the films deposited at 600, 700 and 800 $^0$C respectively. Since
the field for H$_{c\parallel}$ measurement was applied diagonally,
the true value of H$_{c\parallel}$ would be the component of
measured H$_{c\parallel}$ along the [100] or [010] direction. The
correct H$_{c\parallel}$ deduced in this manner then become 74 Oe,
4.2 kOe and 5.9 kOe for the 600, 700 and 800 $^0$C films
respectively. Since all measurements of magnetization in parallel
configuration were performed with field applied along [110], we
will only refer to corrected value of H$_{c\parallel}$ unless
otherwise specified.

The extremely low coercivity (H$_{c\parallel}$ $\approx$ 74 Oe) of
the film deposited at 600 $^0$C is consistent with our X-ray data
which show only the disordered A1 phase at this T$_d$. Such a low
coercivity has been reported by several workers
\cite{Ersen,Park,Xiao,Lewis} in their A1 phase CoPt films. For the
film deposited at 700 $^0$C, the H$_{c\parallel}$ increases to 4.2
kOe. This increase in H$_{c\parallel}$ is also consistent with
X-ray diffraction results which show the formation of the L1$_0$
phase with grains of c-axis in the plane of the substrate as well
as perpendicular to it. Samples prepared at 800 $^0$C display even
higher H$_{c\parallel}$ ($\approx$ 5.9 kOe). Crystallographically,
such films are mostly pure L1$_0$ phase with c-axis normal to the
plane of the substrate.

There are three important features in the M-H$_{\parallel}$ data
of Fig. 1(b) which need to be highlighted; (i) while the
magnetization of the 600 $^0$C film saturates at very low field
($\simeq$ 5 kOe), the film of T$_d$ = 700 and 800 $^0$C show a
non-saturating tendency of magnetization even at the maximum
applied field of $\approx$ 20 kOe, (ii) the total magnetic moment
of the films  goes down with the increasing deposition
temperature, and (iii), the coercivity increases non-monotonically
as the T$_d$ increases from 600 to 800 $^0$C. Since L1$_0$ CoPt
system exhibits very high anisotropy field H$_k$ ($\approx$ 123
kOe) \cite{Weller} with c-axis as the easy-axis of magnetization,
the volume fraction of the film whose c-axis lies perpendicular to
the surface of the substrate would behave more like a paramagnet
for an in-plane measuring field. This would lead to a
non-saturating M-H curve. While the first two observations made on
the magnetization data of Fig. 1(b) indicate that crystallographic
domains of L1$_0$ phase with c-axis in the plane of the substrate
exist in 700 and 800 $^0$C films, their fraction decreases in 800
$^0$C films as suggested by the low remanence and large coercivity
of such films. The 600 $^0$C film on the other hand is
single-phase disordered fcc with its characteristically small
H$_c$.

We have also studied the effects of thermal annealing on crystal
structure and magnetization of a film deposited at 700 $^0$C on
STO. A 30 minutes anneal causes a significant enhancement in the
degree of chemical ordering with c-axis of the ordered phase
aligned parallel to the film normal. The observed increase in
coercivity from 4.2 kOe to 5.9 kOe on annealing is consistent with
the results of structural analysis.

\subsection{Effect of film thickness (t$_F$) and epitaxial strain:}

Fig. 2(a) shows a series of XRD patterns for films (series B) of
thickness ranging from 30 to 100 nm grown on STO at 700 $^0$C and
annealed at the same temperature for 30 minutes. The fundamental
and superlattice peaks of the L1$_0$ phase are observed clearly in
all the samples. We also note that with the increasing thickness,
the intensity of the (200) reflection of L1$_0$ increases and for
100 nm thick film it becomes equal to the intensity of the (002)
peak. The M-H curves for in-plane magnetization M(H$_{\parallel}$)
and the morphology of these films are shown in Fig. 2(b) and Fig.
2(c, d, e) respectively. We observe an interconnected maze-like
pattern of CoPt layer in the 20 nm thick film. On increasing the
thickness to 30 nm, the void fraction in the film decreases, and
at t$_F$ = 50 nm, the percolating network expands at the expense
of the voids and the film becomes almost continuous.

A set of samples was also prepared on MgO substrates, loaded
side-by-side to STO whose results have been discussed in the
preceding section. The XRD profiles, M-H$_{\parallel}$ curves and
surface morphology of these samples are presented in Fig. 3.
Interestingly, the morphological features of these films are
similar to those deposited on STO [Fig. 2] in spite of a large
difference in strain parameters $\epsilon_{MgO}$ and
$\epsilon_{STO}$. It appears that the higher G$_r$ ($\approx$ 1
{\AA}/sec) used here and the moderate growth temperature ($\sim$
700 $^0$C), conspire to increase the supersaturation of adatoms on
the surface of the substrate which reduces the critical size of a
cluster beyond which it does not grow, and increase the cluster
nucleation rate \cite{Greene,Chrisey1}. These two factors appear
to suppress the effect of strain leading to a similar morphology
of the films deposited on STO and MgO.

Fig. 3(a) shows the XRD profiles of the 30, 50 and 100 nm thick
films grown on MgO. These results reveal that unlike the films on
STO, MgO favors a growth with c-axis of the L1$_0$ phase in the
plane of the substrate. Here the intensity of the (200) peak
(I$_{200}$) increases faster than that of the (002) peak
(I$_{002}$). In the inset of Fig. 4 we plot the variation of
I$_{200}$/I$_{002}$ for the films deposited on MgO and STO as a
function of their thickness. It is expected that the increased
fraction of the c-axis in-plane phase on MgO with increasing t$_F$
should lead to a distinct increase in the low-field magnetization
for in-plane measurements. Our data for M(H$_{\parallel}$) (Fig.
4(b)) however, do not reveal a clear trend with t$_F$ within the
accuracy ($\pm$ 15 $\%$) of these measurements. The in-plane
coercivity (H$_{c\parallel}$) of these films as a function of
their thickness is plotted in Fig. 4. A marginal enhancement in
coercivity is seen as the films are made thinner. By comparing
this result with the microstructure of the films, it becomes clear
that a percolative network is more coercive than a continuous
film.

In order to understand the magnetization reversal process in films
of ordered L1$_0$ phase with mixed crystallographic domains, some
with their c-axis parallel and other perpendicular to the
substrate plane, we have measured the magnetization of the series
`B' samples [T$_d$ = 700 $^0$C] grown on STO in H$_{\parallel}$
and H$_{\perp}$ geometries. The magnetic field for the in-plane
(H$_{\parallel}$) measurements was applied at 45$^0$ with respect
to the [100] and [010] direction of the substrate as mentioned in
the previous section. The component of H$_{c\parallel}$ along
[100] or [010] direction is expected to be equal to H$_{c\perp}$
if the three variants of c-axis are present with equal abundance.
Interestingly, the calculated values of the H$_{c\parallel}$ along
[100] and [010] (listed in Table -II) match reasonably well with
the measured values of H$_{c\perp}$ irrespective of the film
thickness.

\subsection{Fractals and nano-dots at low growth rate:}
The evolution of film morphology with deposition temperature and
growth rate is intimately linked with the epitaxial strain
parameter $\epsilon$, provided the growth rate is not large
\cite{Kim,Budhani}. In order to study the strain induced
crystallographic and morphological changes, a set of samples
(series `C') was prepared using a slow deposition rate ($\approx$
0.4 {\AA}/sec) on (001) LAO, (001) STO and (001) MgO. From the XRD
data we could not conclude whether there exist any c-axis variants
in the film on LAO when grown at 700 $^0$C because the position of
the (200) reflection in $\theta$ - 2$\theta$ scan is same as of
the (002) (using pseudocubic unit cell) reflection of the
substrate. But there is a clear signature of this in films
deposited on the other two substrates. It was also evident from
the relative intensity of the (200) and (002) reflections that the
fraction of the domains with in-plane c-axis is more in the film
on MgO as compared to that in the film on STO. The c-axis lattice
parameter of these films calculated from the position of the (002)
reflection shows a small but distinct drop at the higher growth
temperatures (see Fig. 5). This change is independent of the
substrate used. At the highest deposition temperature i.e., 800
$^{0}$C, the c-axis lattice parameter for films on all the three
substrates compare reasonably well with the bulk value of 3.701
{\AA}.

The SEM micrographs of the surface of these films present some
interesting manifestations of epitaxial strain. As seen in Fig.
6(a) for the film deposited on STO at 700 $^0$C, the morphology is
a self-similar fractal where the fractals do not form a
percolating backbone over a length scale larger than $\simeq$ 5
$\mu$m. On increasing the T$_d$ to 750 $^0$C, the maze-like
structure breaks up into islands. A similar morphology is seen in
films deposited on MgO at 750 $^0$C. Upon further increase of
T$_d$ to 800 $^0$C, the islands on STO acquire a smaller and
nearly uniform size distribution ($\simeq$ 200 nm) [Fig. 6(c)].
The values of the thermal expansion coefficients of STO and MgO
(11.1 $\times$ 10$^{-6}$ K$^{-1}$ and 12.8 $\times$ 10$^{-6}$
K$^{-1}$, respectively \cite{Hwang}) are close to that of CoPt
(10.9 $\times$ 10$^{-6}$ K$^{-1}$ \cite{Liao}). These data suggest
that the $\epsilon$ parameter at T$_d$ does not differ much from
its room temperature value for MgO as well as STO. The breaking of
the films can, therefore, be understood on the basis of the strain
parameter calculated using room temperature lattice constants. The
reason for these changes in film morphology can be argued to
depend on the imbalance of interfacial and strain energies that
minimizes the contact area between film and the substrate with
increasing T$_d$ \cite{Kim}.

The most interesting case is when the films are grown on LAO which
has nearly perfect lattice match ($\epsilon$ = + 0.3 $\%$) with
the out-of-plane L1$_0$ CoPt at room temperature. As evident from
Fig. 7 there is an irregular shaped island growth on LAO even at
700 $^0$C. Since the coefficient of thermal expansion for LAO (9.2
$\times$ 10$^{-6}$ K$^{-1}$) is similar to that of CoPt
\cite{Hwang}, this breaking of the film can not be explained on
the basis of strain alone. However, LAO undergoes a structural
phase change from rhombohedral to cubic symmetry on heating above
800 $^0$K \cite{Qinxin}. The lattice parameter and associated
expansion coefficient of the cubic phase perhaps facilitate the
island growth seen at 700 $^0$C. The island morphology becomes
sharper on increasing the T$_d$ to 800 $^0$C. This nanoscale
structure that evolves at the higher T$_d$ and low G$_r$ is likely
to be due to the enhanced desorption of adatoms from surface which
leave large areas with little or no coverage. These regions can
host new islands \cite{Jensen}. Similar observations have been
made by Castaldi et al \cite{Castaldi} for CoPt films deposited on
oxidized Si substrates using electron beam evaporation technique.

\subsection{Magnetic behavior of the fractals and nano-dots grown on STO:}
We have addressed the magnetization and demagnetization
characteristics of the films deposited at low growth rate (G$_r$
$\approx$ 0.4 {\AA}/sec) on STO (series C) in detail. The typical
M-H loops at 300 K of the samples grown at 700, 750 and 800 $^0$C
are shown in Fig. 6 (d, e and f) respectively. These data are for
field applied perpendicular to the plane of the film (
H$_{\perp}$), except for 800 $^0$C sample for which
H$_{\parallel}$ measurement is also shown. A striking feature of
these data is a substantial increase in coercivity (H$_{c\perp}$)
[see Table - II] with the increasing deposition temperatures. In
order to understand the magnetization dynamics and its reversal
mechanism let us consider the three samples separately.

Sample deposited at 700 $^0$C does not show any saturation of the
magnetic moment even at 30 kOe. The coercive field H$_{c\perp}$ in
this case is $\approx$ 7 kOe. We have already discussed the
presence of c-axis variants of L1$_0$ phase with grains of c-axis
parallel to [100] and [010] directions of the substrate for the
sample deposited at 700 $^0$C. Since c-axis is the easy-axis of
magnetization in L1$_0$ phase of CoPt film, the fraction of L1$_0$
phase with in-plane c-axis play a major role in setting the
magnetic response of the sample. The paramagnet-like contribution
to magnetization of the in-plane c-axis fraction of CoPt when the
field is applied perpendicular to substrate surface would display
a non-saturating tendency of magnetic moment even at very high
fields because of the large uniaxial anisotropy (H$_k$ $\approx$
123 kOe) of L1$_0$ CoPt phase. The remanence squareness S (=
M$_{r}$/M$_{s}$), where M$_{s}$ and M$_r$ are the saturation and
remanent magnetization respectively, of the series `C' samples is
plotted in Fig. 8 as a function of T$_{d}$. The S parameter is
$\simeq$ 0.85 for the 700 $^0$C deposited films. Although a high
value of S ($>$ 0.5) has been attributed to exchange coupling
amongst the particles \cite{Hao}, we believe that the large S seen
in this sample of a fractal-like morphology is a manifestation of
pinning of the domain walls by the boundaries between adjacent
c-axis variants where the easy axis changes by 90 $^0$
\cite{Lewis}.

For the sample deposited at 800 $^{0}$C the saturation field is
beyond the maximum dc-field (46 kOe) used in these M (H$_{\perp}$)
measurements. This is indicated by the hysteresis loop shown in
Fig. 6(f), which is neither symmetrical nor centered about the
origin, and has no reversible region even at the maximum applied
field. A very large H$_{c\perp}$ of $\approx$ 30 kOe is seen in
these films. It is important to mention here that this
H$_{c\perp}$ has been derived by taking the average of the
magnetic field corresponding to zero moment from both sides of the
hysteresis loop. The true H$_{c\perp}$ may be still higher because
the loop did not reach saturation due to the limited value of the
applied field \cite{Jacobs}. Fig. 6(f) also shows the M-H curve
taken in the parallel field configuration. The uniaxial magnetic
anisotropy K$_u$ of this film, determined from the area enclosed
by the parallel and perpendicular magnetization curves, comes out
to be $\approx$ 4.62 $\times$ 10$^{6}$ J/m$^3$, which is very
close to the maximum value ($\sim$ 5.0 $\times$ 10$^{6}$ J/m$^3$)
reported in the literature \cite{McCurrie}.

In order to establish a correlation between morphological features
and magnetic domain structure of the films of the series `C', we
show in Fig. 9 the atomic and magnetic force micrographs (AFM and
MFM respectively) taken from the same spot on the film. For the
750 $^0$C sample, there is some discrepancy between the SEM image
(Fig. 6(b)) and the AFM image (Fig. 9(a)). While the SEM image
shows well-separated grains, the AFM image reveals a much smaller
separation between them. The MFM image of this sample suggests
that the magnetic domains are not confined to a physical grain,
and also the boundary between two adjacent domains is not sharp.
This clear non-conformity of the topographical structure and
magnetic domain pattern suggest that the grain boundaries do not
play a significant role in pinning of domains and their reversal.
The AFM image of the sample grown at 800 $^0$C (Fig. 9(b)) is
quite similar to its SEM image (Fig. 6(c)) which shows uniform and
well-separated nano grains. The MFM image of individual grains
does not show any domain boundaries. Since one can exactly map
each feature of the AFM image onto the MFM image, it is clear that
each grain in the film is a single magnetic domain. The bright and
dark features in the MFM image, which represent nano grains of
reversed magnetization are similar in number, ensuring zero
magnetization in the thermally demagnetized state. Another
interesting aspect of this image is that the bright and dark
features tend to group among themselves showing string-like
structures and islands. Although the remanence squareness
parameter S ($\simeq$ 0.59) for this sample is close to that for a
randomly oriented assembly of non-interacting Stoner-Wohlfarth
(S-W) particles \cite{Hao, Stoner}, we cannot really rule out
interaction between the particles. Further, a limited spatial
resolution of MFM ($\approx$ 50 nm) also makes it difficult to say
whether the demagnetization process is via a coherent rotation of
magnetization or due to a curling mode which can lead to vorticity
in the spin system.

\subsection{Relaxation of magnetization in single-domain nano-dots:}
In order to understand the process of magnetization reversal,
relaxation measurements of magnetization at constant negative bias
field close to the coercive value (H$_{c\perp}$) were carried out
on film deposited at 800 $^{0}$C (series `C'). The theory of
thermal fluctuation of the magnetic moment of single-domain
ferromagnetic particles was introduced by N\'{e}el \cite{Neel} and
further developed by Brown \cite{Brown}. For the particles at zero
applied field, the energy barrier between two equilibrium states
of opposite moment is too high to observe magnetization reversal
in a reasonable experimental time scale of a few hours. However,
by applying a reverse field close to the coercive field H$_{c}$
after fully magnetizing the particle, reversal can be initiated
through thermal activation process within a short time scale. In
our relaxation measurements, the sample was first magnetized at
300 K in a field of 46 kOe applied perpendicular to its plane. The
field was then reversed to a desired value H ($\leq$ H$_{c\perp}$)
in one step, and sample moment was monitored as a function of time
over a period of 2.5 hours. Fig. 10 shows the decay of
magnetization as a function of time at different values of
reversed field. The presence of two distinct regimes of time in
the figure indicates that there are two different activation
processes leading to the relaxation of the system. A distribution
in grain size is a possible reason for this behavior. The data of
Fig. 10 have been fitted to the N\'{e}el-Brown model
\cite{Wohlfarth} for relaxation given as M(t) = M$_{0}$
exp(-t/$\tau$), where M$_{0}$ is the magnetization at t = 0 and
M(t) the remanent magnetization at time t after removal of the
field, $\tau$ is the relaxation time defined as 1/$\tau$ = f$_{0}$
exp(- E(H)/k$_{\beta}$T), f$_{0}$ is a frequency factor of the
order of 10$^{9}$ sec$^{-1}$ and E(H) = K$_{u}$V(1 -
H/H$_{k}$)$^{2}$ the energy barrier in a reverse field of
magnitude H. Here V, K$_u$ and H$_k$ denote particle volume,
anisotropy energy density and anisotropy field (=2K$_u$/M$_s$)
respectively \cite{Dormann}. It appears that the relaxation rate
of magnetization increases rapidly as the absolute value of the
magnetic field is increased. Using the above formula and
approximating the grains to be of a spherical shape, we extract
the activation volume whose width is $\approx$ 4.7 nm. This is far
less than the diameter ($\approx$ 200 nm) of the disc-shaped
magnetic particles seen in SEM and MFM micrographs. This
discrepancy points towards inadequacy of the model used.

The relaxation behavior of an assembly of single-domain particles
can deviate from the ideal N\'{e}el - Brown picture due to several
reasons which include, multi-domain nature of particles,
interaction between the particles and a distribution in their
size. Our MFM images, however, do not show any well-defined
sub-domains in the nano grains. Interaction amongst particles, on
the other hand, leads to an apparent enhancement rather than a
decrease in the activation volume. One could argue that the
demagnetization process occurs via a curling mode in which the
magnetization continuously curls around the particle center
\cite{Zijlstra}. This can, in principle, lead to a smaller
activation energy for demagnetization. A simple calculation for
our disc-shaped grains of diameter $\approx$ 200 nm and thickness
$\approx$ 50 nm with anisotropy axis perpendicular to the disc
plane shows that reversed spins on a circular ring of width
$\approx$ 60 nm around the periphery of the disc would be
sufficient to make the net magnetization zero. However, the
presence of such a curing mode can not be established with MFM due
to its limited resolution. Further work, preferably with spin
polarized tunneling, as done in the case of Fe platelets grown on
(110) surface of tungsten \cite{Wachowiak} may shed more light on
the demagnetization dynamics of these CoPt nano-dots.

\section{Conclusions}
CoPt films of 20 to 100 nm thickness were successfully grown using
the Pulsed Laser Deposition (PLD) technique by ablating a
Co$_{50}$Pt$_{50}$ alloy target. The deposition conditions which
facilitate growth of the L1$_0$ ordered phase of high
magnetocrystalline anisotropy are clearly identified. Correlations
between film morphology, crystallographic structure and magnetic
properties such as magnetocrystalline anisotropy, shape and size
of the hysteresis loops, coercive field and magnetic domain
structure etc. have been established. We note that the
crystallographic ordering in films deposited at 600 $^0$C on
single crystal substrates which offer different degree of strain
is similar. These films have a disordered fcc structure with
in-plane magnetization and a narrow hysteresis loop of width
$\approx$ 200 Oe. The near-complete chemical ordering achieved at
T$_d$ $\geq$ 700 $^0$C, becomes predominantly c-axis oriented as
the deposition temperature is raised to 800 $^0$C. While the
changes in the crystallographic structure of the films depend
solely on T$_d$, their microstructure and magnetization
characteristics are decided by the growth rate. It is observed
that irrespective of a large difference in lattice parameters of
the substrates used, there is hardly any change in morphology of
the films grown on different substrates, when a high growth rate
($\approx$ 1 {\AA}/sec) is maintained. The microstructure of these
L1$_0$ films evolves from a well-connected maze-like structure at
t$_F$ = 20 nm to a smooth and continuous film at t$_F$ = 50 nm.
The magnetic coercivity of these films increases as the L1$_0$
phase fraction grows with T$_d$ and its orientation becomes out of
the film plane.

While the higher growth rate seems to suppress the effects of
strain due to lattice mismatch, low growth rate ($\approx$ 0.4
{\AA}/sec) reveals granularity in the films even at small
thickness ($\approx$ 50 nm). Here the structure changes from a
self-similar fractal pattern to nano-dots as the T$_d$ is raised
from 700 to 800 $^0$C. The reason for this topographical change
appears to be the imbalance between strain and interfacial
energies, which is overridden by kinetic considerations at the
high growth rates. The AFM and MFM images of the films grown at
low G$_r$ and high T$_d$ reveal physically separated, magnetically
single-domain nano-scale islands. Magnetic relaxation measurements
carried out to study the relaxation dynamics of these
single-domain islands show negligible decay of magnetization
unless a reverse field close to the coercive field (H$_c$
$\approx$ 30 kOe) is applied. The simple picture of coherent
rotation of magnetization appears incompatible with the time
dependence of the remanent magnetization.

This research has been supported by a grant from the National
Nanoscience $\&$ Nanotechnology Initiative of the Department of
Science $\&$ Technology, India.

\clearpage

\begin{table}
\caption{Strain parameter $\epsilon$ for the growth of fcc and
L1$_0$ CoPt on different substrates. The lattice parameter of LAO,
MgO and STO substrates are 3.792 {\AA}, 4.21 {\AA} and 3.905 {\AA}
respectively. While CoPt in fcc phase is cubic with lattice
parameter of 3.772 {\AA}, the lattice parameters of the L1$_0$
CoPt in the bulk form are a = b = 3.803 {\AA} and c = 3.701
{\AA}.}
\begin{tabular}{ccccccccccccc}
\hline \hline
substrate &&&& crystallographic &&&& growth &&&& $\epsilon$\\
&&&& structure of CoPt &&&& direction &&&&\\
\hline

LAO (001) &&&& fcc &&&& -- &&&& - 0.5 $\%$\\
LAO (001) &&&& L1$_0$ &&&& in-plane c-axis &&&& - 2.5 $\%$\\
LAO (001) &&&& L1$_0$ &&&& out-of-plane c-axis &&&& + 0.3 $\%$\\
\hline
MgO (001) &&&& fcc &&&& -- &&&& - 11.6 $\%$\\
MgO (001) &&&& L1$_0$ &&&& in-plane c-axis &&&& - 13.8 $\%$\\
MgO (001) &&&& L1$_0$ &&&& out-of-plane c-axis &&&& - 10.7 $\%$\\
\hline
STO (001) &&&& fcc &&&& -- &&&& - 3.5 $\%$\\
STO (001) &&&& L1$_0$ &&&& in-plane c-axis &&&& - 5.5 $\%$\\
STO (001) &&&& L1$_0$ &&&& out-of-plane c-axis &&&& - 2.7 $\%$\\
\hline \hline

\end{tabular}
\label{table1}
\end{table}

\clearpage

\begin{table}
\caption{Sample preparation conditions for different series of
films. The table also lists the in-plane and out-of-plane coercive
field of the films.}
\begin{tabular}{cccccccccccccccccccccccc}
\hline \hline
series  &&& G$_r$       &&& t$_F$ &&& T$_d$  &&&& substrate &&&& post growth annealing &&& H$_{c\parallel}$/H$_{c\perp}$ \\
        &&& ({$\approx$ \AA}/sec) &&& (nm)  &&& ($^0$C)&&&&           &&&& time (minutes) &&& (kOe)\\
\hline
A  &&& 1 &&& 100 &&& 600 &&&& STO &&&& No &&& 0.074/-- \\
   &&& 1 &&& 100 &&& 700 &&&& STO &&&& No &&& 4.2/-- \\
   &&& 1 &&& 100 &&& 800 &&&& STO &&&& No &&& 6/-- \\
   &&& 1 &&& 100 &&& 700 &&&& STO &&&& 30 &&& 5.9/-- \\
\hline
B  &&& 1 &&& 20 &&& 700 &&&& STO &&&& 30  &&& --/-- \\
   &&& 1 &&& 30 &&& 700 &&&& STO &&&& 30  &&& 6.4/6.7 \\
   &&& 1 &&& 50 &&& 700 &&&& STO &&&& 30  &&& 6.2/5.1 \\
   &&& 1 &&& 100 &&& 700 &&&& STO &&&& 30 &&& 5.4/4.3 \\
   &&& 1 &&& 20 &&& 700 &&&& MgO &&&& 30  &&& --/-- \\
   &&& 1 &&& 30 &&& 700 &&&& MgO &&&& 30  &&& 8.3/--\\
   &&& 1 &&& 50 &&& 700 &&&& MgO &&&& 30  &&& 7.9/-- \\
   &&& 1 &&& 100 &&& 700 &&&& MgO &&&& 30 &&& 6.8/--\\
\hline
C  &&& 0.4 &&& 50 &&& 700 &&&& LAO &&&& 25 &&& --/-- \\
   &&& 0.4 &&& 50 &&& 750 &&&& LAO &&&& 25 &&& --/--\\
   &&& 0.4 &&& 50 &&& 800 &&&& LAO &&&& 25 &&& --/-- \\
   &&& 0.4 &&& 50 &&& 700 &&&& MgO &&&& 25 &&& --/-- \\
   &&& 0.4 &&& 50 &&& 750 &&&& MgO &&&& 25 &&& --/--\\
   &&& 0.4 &&& 50 &&& 800 &&&& MgO &&&& 25 &&& --/--\\
   &&& 0.4 &&& 50 &&& 700 &&&& STO &&&& 25 &&& --/7 \\
   &&& 0.4 &&& 50 &&& 750 &&&& STO &&&& 25 &&& --/16\\
   &&& 0.4 &&& 50 &&& 800 &&&& STO &&&& 25 &&& 0.4/37\\
\hline \hline

\end{tabular}
\label{table2}
\end{table}

\clearpage

\section*{FIGURE CAPTIONS:}

\noindent FIG. 1. (a) X-ray diffraction profiles of the 100 nm
films deposited at various temperatures on (001) cut STO. The
diffraction profile of a bare substrate is also shown (bottom
panel). Arrows in the top panel mark the position of fundamental
as well as superlattice reflections of the L1$_0$ ordered phase.
(b) In-plane magnetization of the same set of samples whose XRD is
shown in Fig. 1(a). A sketch of the direction of in-plane field
H$_{\parallel}$ is shown in the lower inset of the figure. The
upper inset represents a schematic of the three c-axis variants of
the L1$_0$ phase.

\noindent FIG. 2. (a) X-ray diffraction profiles of CoPt thin
films of various thickness deposited at 700 $^0$C on STO (001) at
a growth rate of $\approx$ 1 {\AA}/sec. All the samples were
annealed for 30 minutes at the growth temperature. The diffraction
profile of single crystal STO (001) is also shown (bottom panel).
Arrows in Fig. 2(a) indicate the position of fundamental as well
as superlattice reflections of the L1$_0$ phase. Dotted line in
the figure marks the position of the (200) reflection. In-plane
magnetization curves of the same set of samples are shown in panel
(b). Panels (c), (d) and (e) show SEM micrographs of the samples
of different thickness.

\noindent FIG. 3. (a) X-ray diffraction profiles of CoPt thin
films of various thickness deposited at 700 $^0$C on MgO (001) at
a growth rate of $\approx$ 1 {\AA}/sec. All the samples were
annealed for 30 minutes at the growth temperature. The diffraction
profile of single crystal MgO (001) is also shown (bottom panel).
Arrows in the figure indicate the position of fundamental as well
as superlattice reflections of the L1$_0$ phase. Dotted line in
the figure marks the position of the (200) reflection. Panel (b)
shows the in-plane magnetization of the same set of samples.
Panels (c), (d) and (e) show SEM micrographs of the samples of
different thickness.

\noindent FIG. 4. Variation of the in-plane coercive field
H$_{c\parallel}$ with thickness for the films on STO and MgO
deposited at 700 $^0$C at a growth rate of $\approx$ 1 {\AA}/sec.
Inset shows the dependence of I$_{200}$/I$_{002}$ on film
thickness. All the samples were annealed for 30 minutes at the
growth temperature. Solid lines in the figure are guide to the
eye.

\noindent FIG. 5. The c-axis lattice parameter plotted as a
function of the deposition temperature of the films on LAO, STO
and MgO at a rate of $\approx$ 0.4 {\AA}/sec. All the samples were
annealed for 25 minutes at growth temperature.

\noindent FIG. 6. SEM images and perpendicular magnetization loops
of 50 nm CoPt thin films deposited at various substrate
temperatures at a growth rate of $\approx$ 0.4 {\AA}/sec on single
crystal STO (001). All the samples were annealed for 25 minutes.
Magnetization panel (f) for the sample grown 800 $^0$C also shows
data for in-plane configuration of the magnetic field.

\noindent FIG. 7. SEM images of of 50 nm CoPt thin films deposited
at 700 and 800 $^0$C with a growth rate of $\approx$ 0.4 {\AA}/sec
on single crystal LAO (001). These samples were annealed for 25
minutes at the growth temperature.

\noindent FIG. 8. Variation of remanent ratio S and perpendicular
coercive field H$_{c\perp}$ with deposition temperature for 50 nm
CoPt thin films deposited at a growth rate of $\approx$ 0.4
{\AA}/sec on single crystal STO (001). All the samples were
annealed for 25 minutes at the growth temperature.

\noindent FIG. 9. (color online). AFM and MFM images of 50 nm CoPt
thin films deposited at (a) 750 $^0$C and (b) 800 $^0$C at a
growth rate of $\approx$ 0.4 {\AA}/sec on STO (001). Measurements
were carried out on thermally demagnetized samples.

\noindent FIG. 10. Magnetic relaxation data collected at T =300 K
for different values of the bias field. The sample was initially
magnetized at 46 kOe. Solid lines are fit to the N\'{e}el-Brown
equation for relaxation \cite{Wohlfarth}.

\clearpage

\begin{figure}[h]
\vskip 0cm \abovecaptionskip 0cm
\includegraphics [width=12cm]{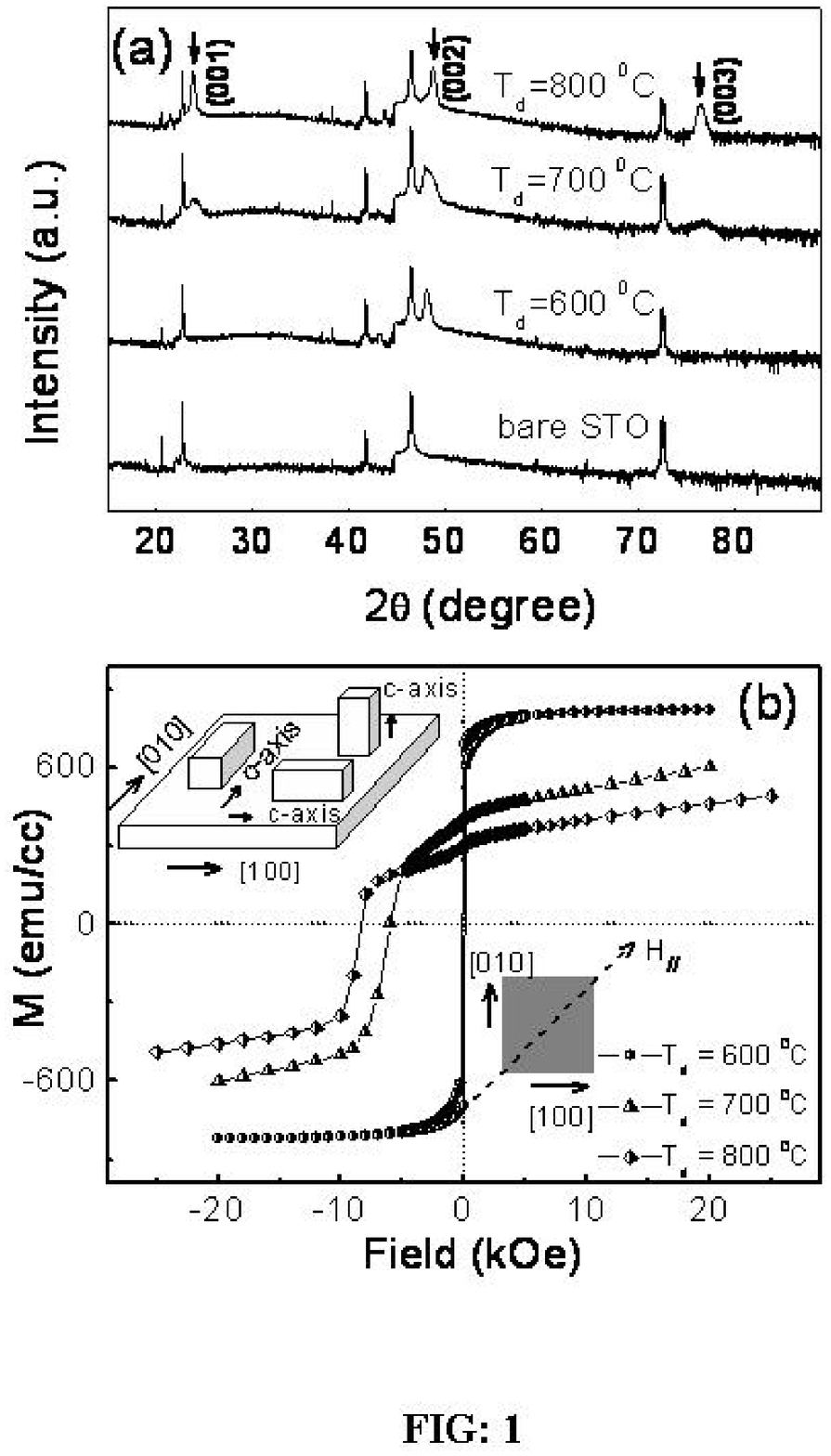}%
\end{figure}

\clearpage
\begin{figure}[h]
\vskip 0cm \abovecaptionskip 0cm
\includegraphics [width=16cm]{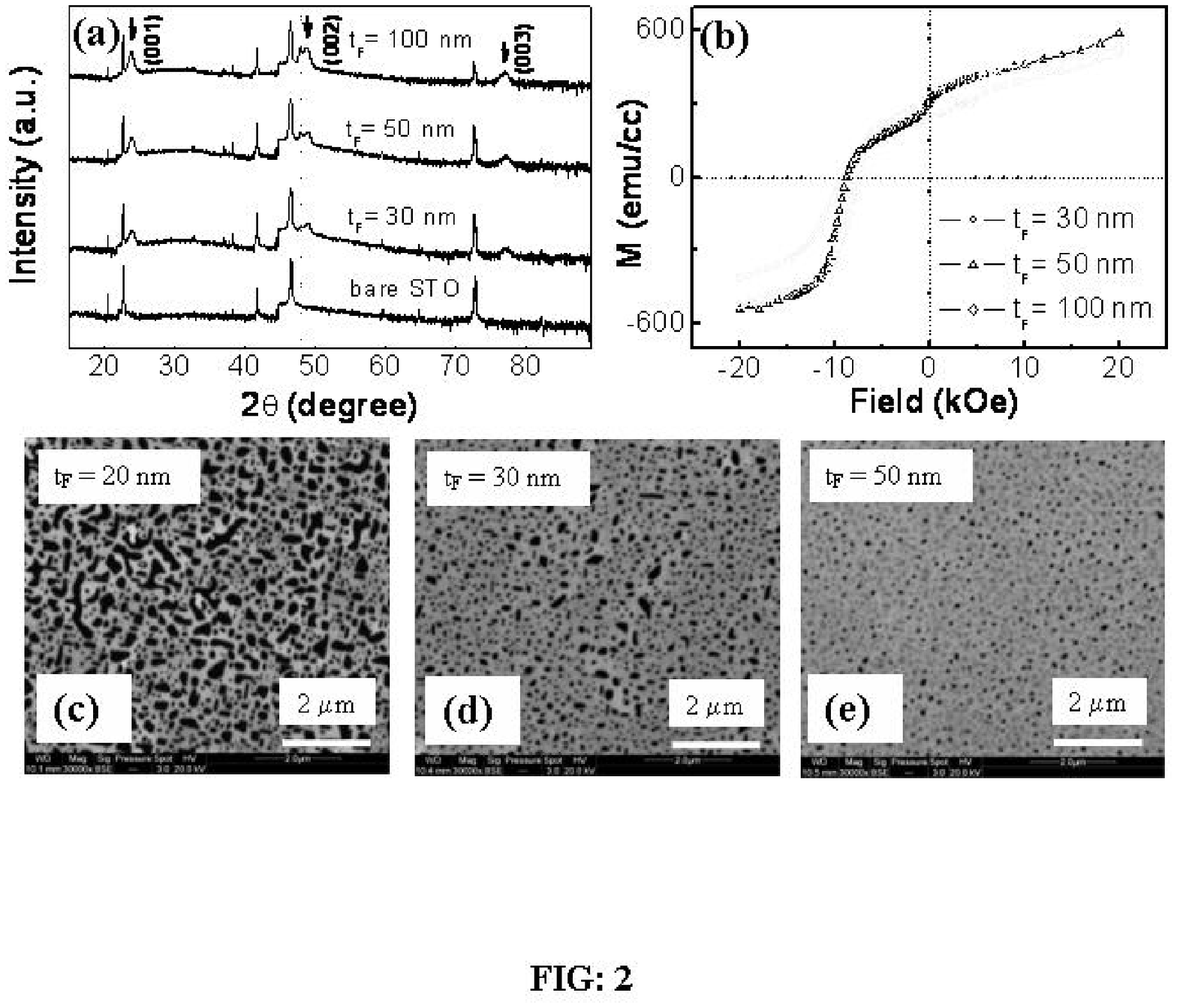}%
\end{figure}

\clearpage
\begin{figure}[h]
\vskip 0cm \abovecaptionskip 0cm
\includegraphics [width=16cm]{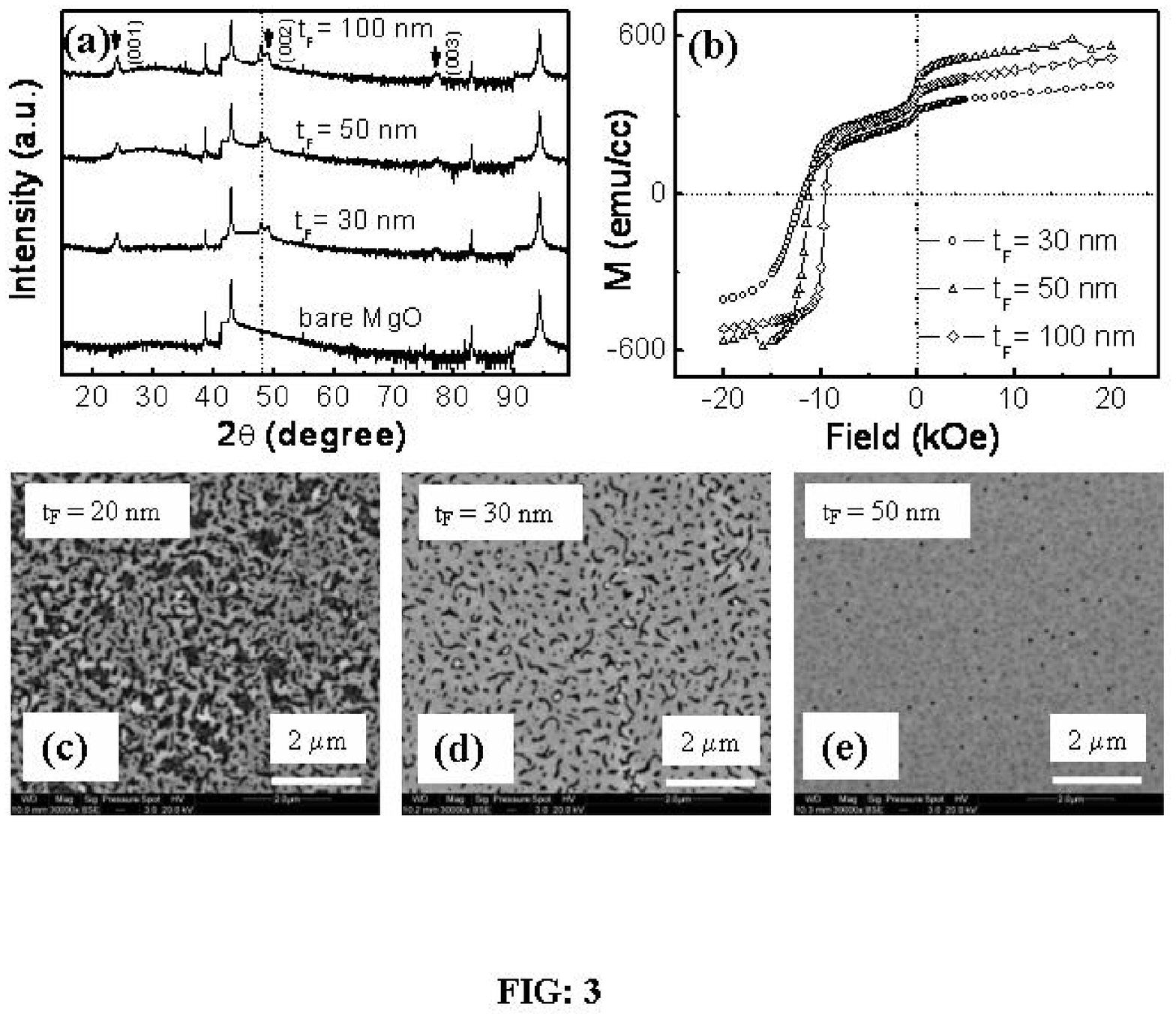}%
\end{figure}

\clearpage
\begin{figure}[h]
\vskip 0cm \abovecaptionskip 0cm
\includegraphics [width=12cm]{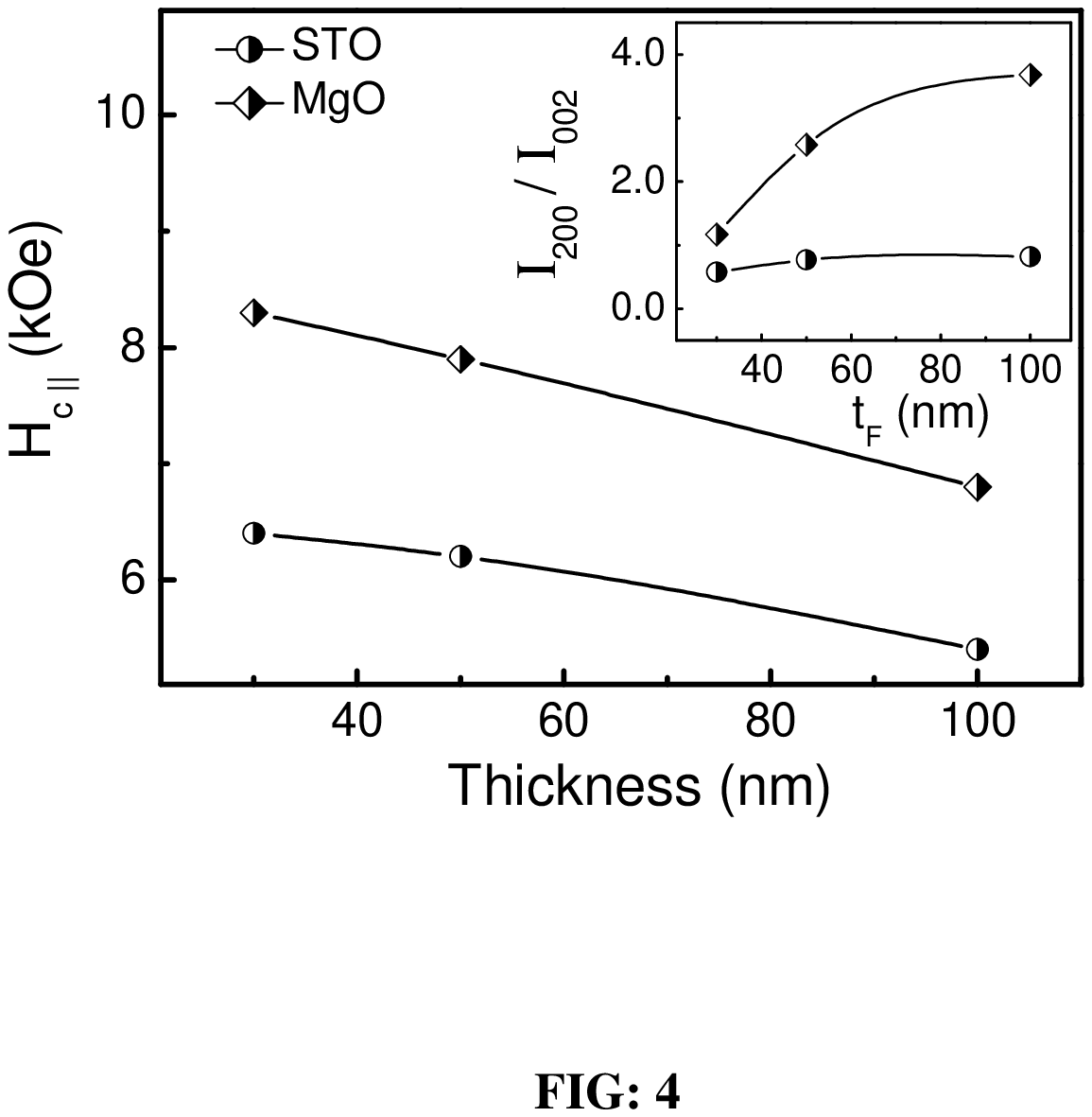}%
\end{figure}

\clearpage
\begin{figure}[h]
\vskip 0cm \abovecaptionskip 0cm
\includegraphics [width=10cm]{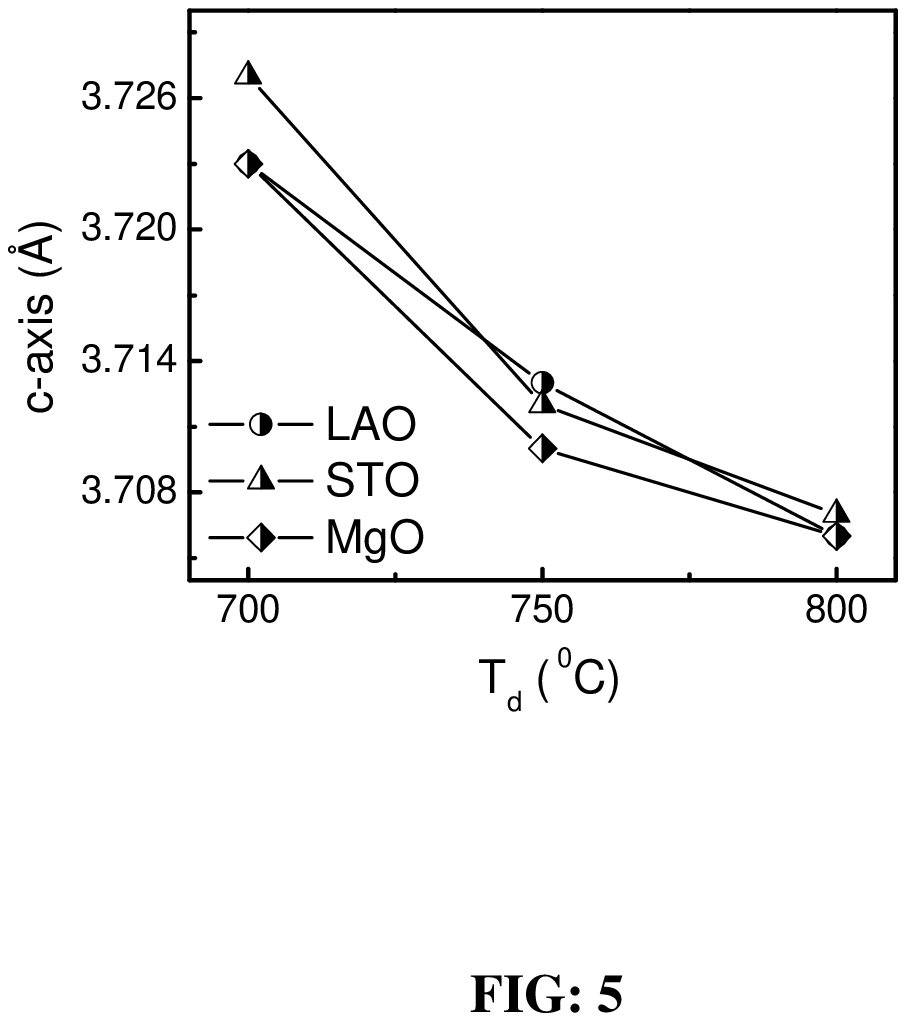}%
\end{figure}

\clearpage
\begin{figure}[h]
\vskip 0cm \abovecaptionskip 0cm
\includegraphics [width=14cm]{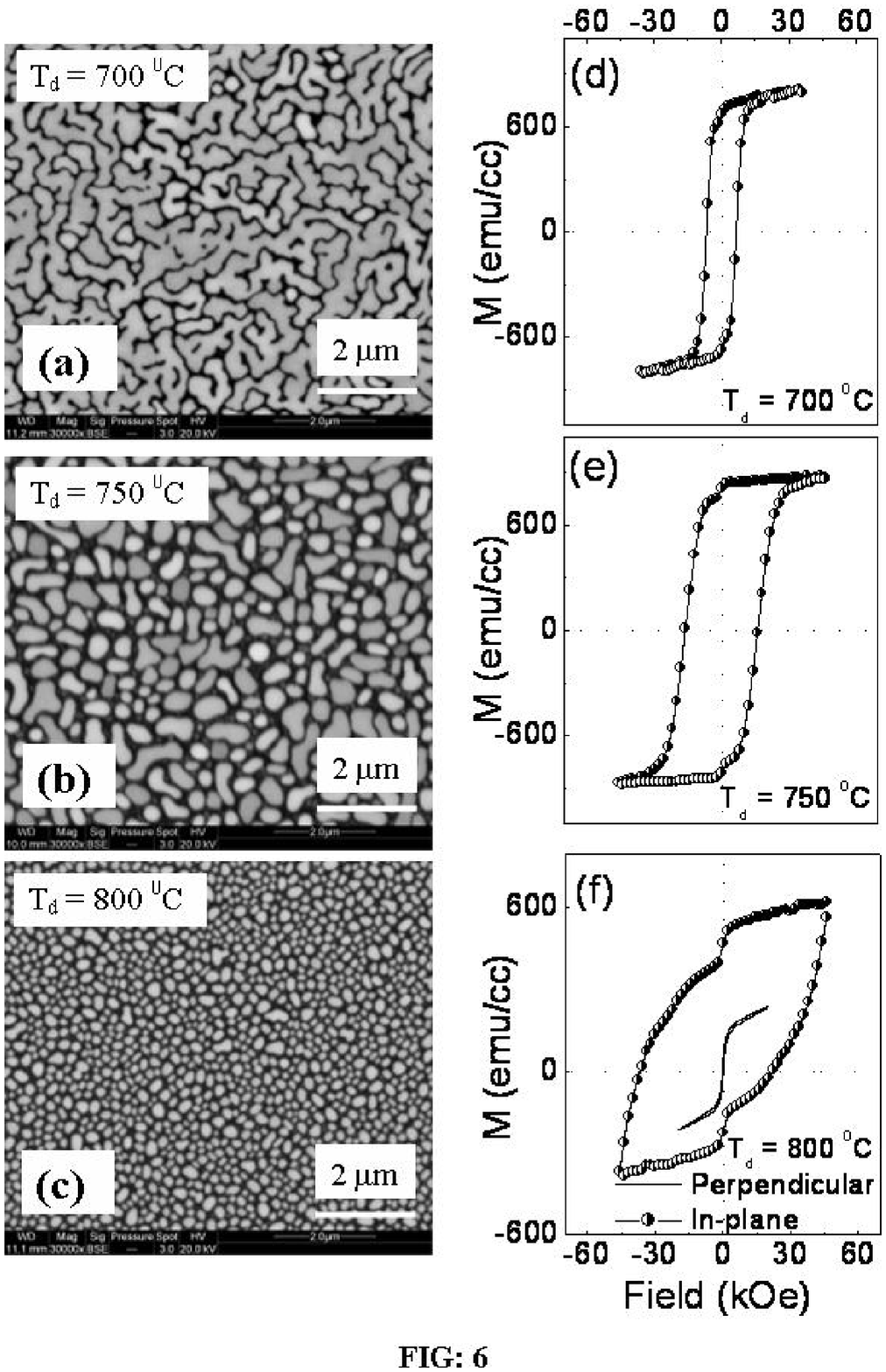}%
\end{figure}

\clearpage
\begin{figure}[h]
\vskip 0cm \abovecaptionskip 0cm
\includegraphics [width=13cm]{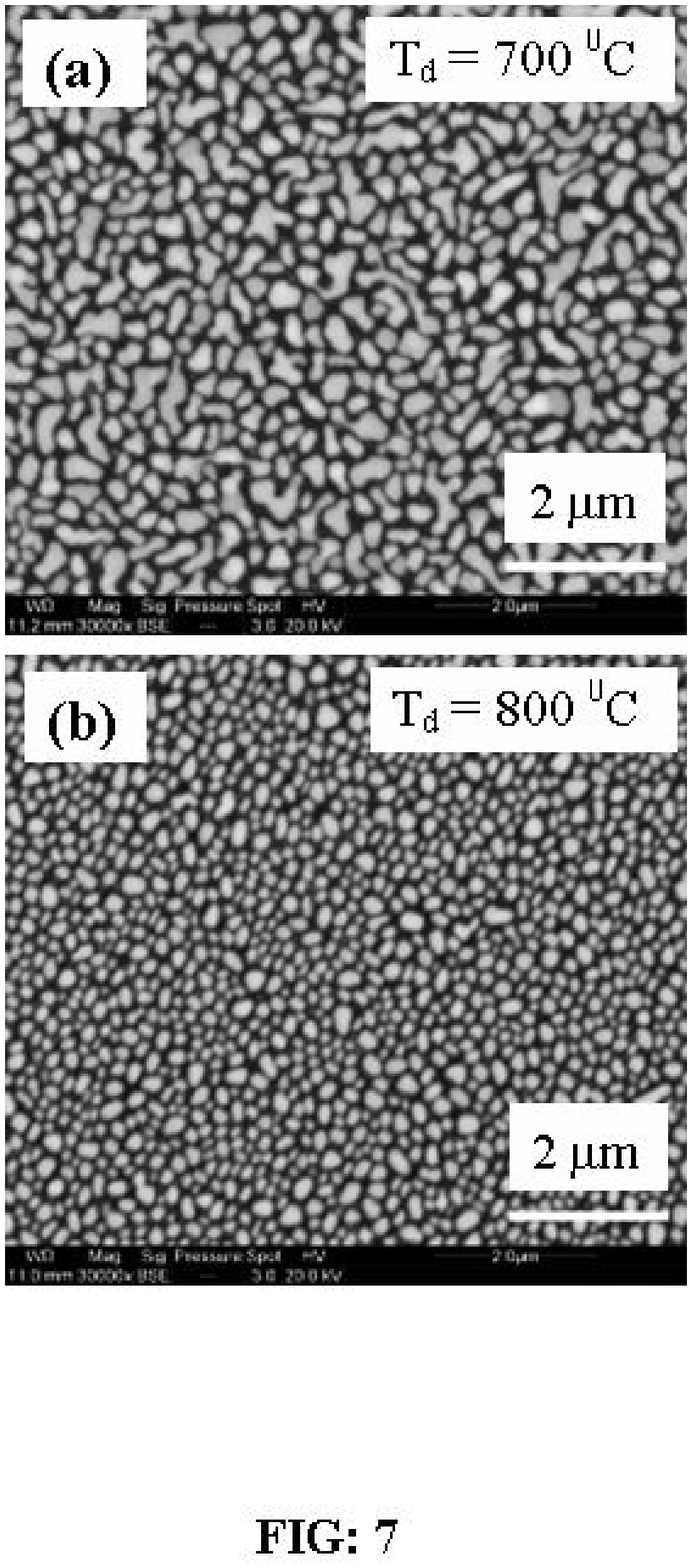}%
\end{figure}

\clearpage
\begin{figure}[h]
\vskip 0cm \abovecaptionskip 0cm
\includegraphics [width=14cm]{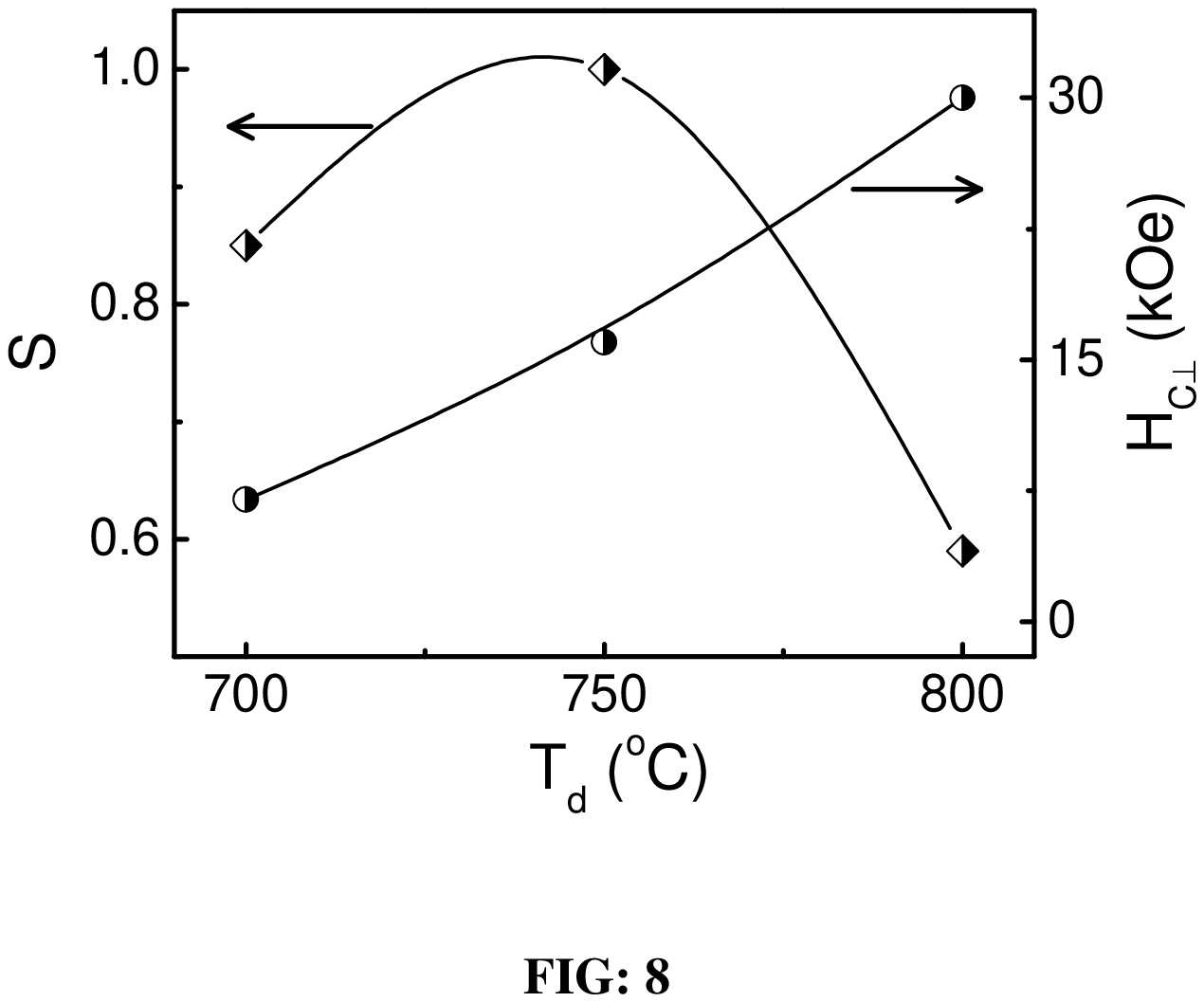}%
\end{figure}
\clearpage

\begin{figure}[h]
\vskip 0cm \abovecaptionskip 0cm
\includegraphics [width=14cm]{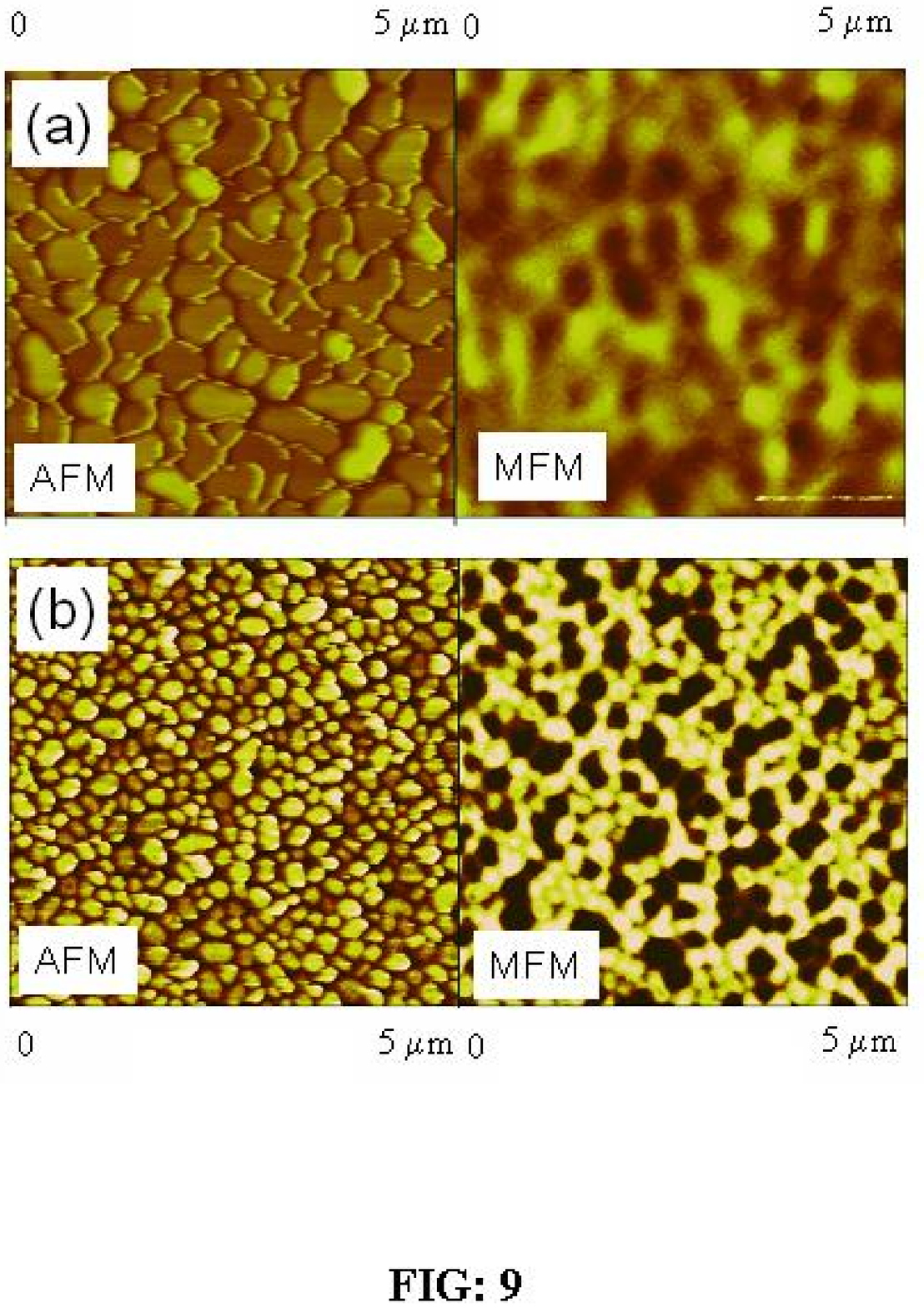}%
\end{figure}

\clearpage
\begin{figure}[h]
\vskip 0cm \abovecaptionskip 0cm
\includegraphics [width=12cm]{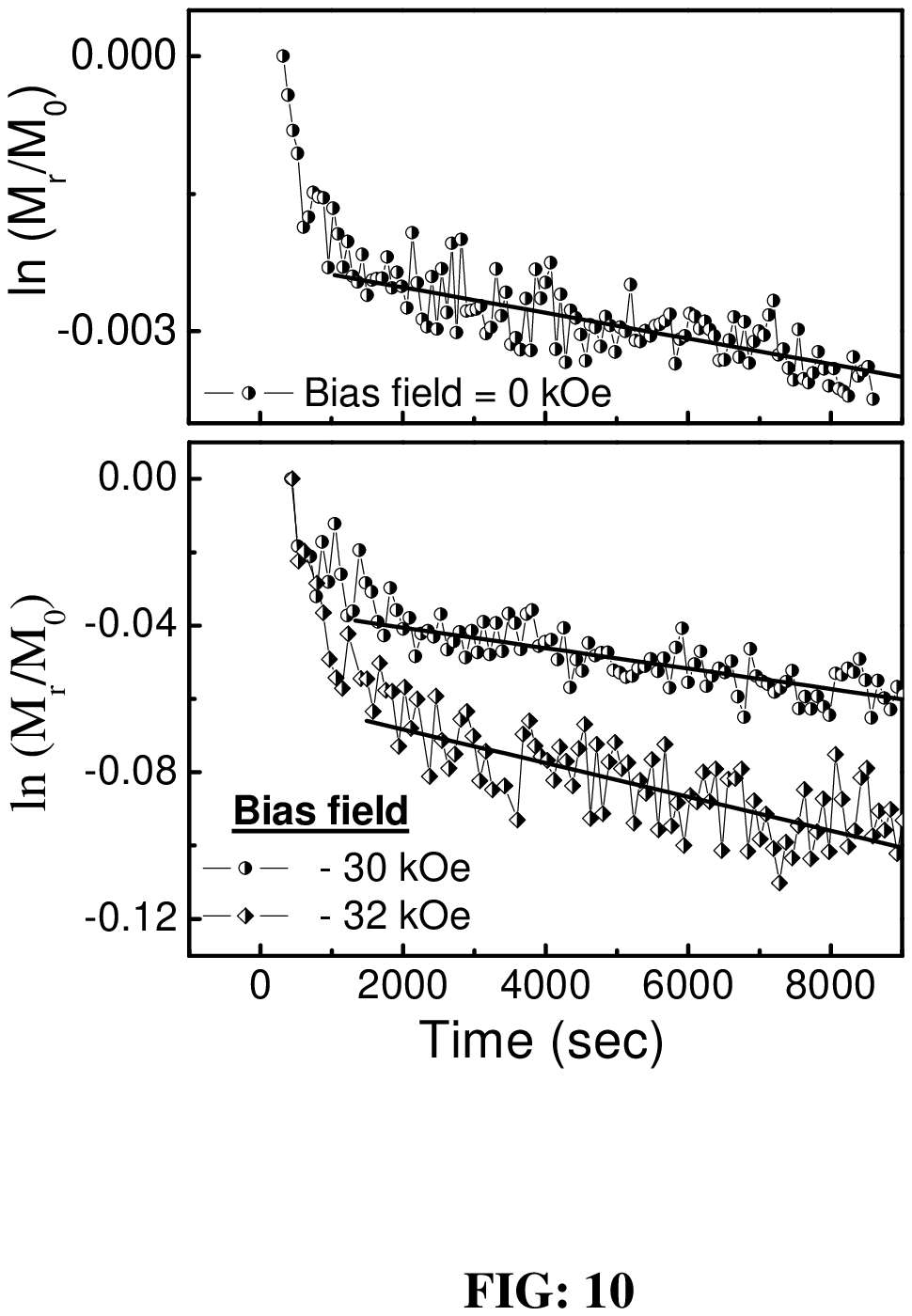}%
\end{figure}

\end{document}